\documentclass[11pt]{article}
\usepackage{amsmath,amsxtra}
\usepackage{amscd}
\usepackage[dvips]{graphicx}
\usepackage{graphicx}
\usepackage{latexsym}
\usepackage{dsfont}

\usepackage{amssymb}
\usepackage{latexsym}
\usepackage{amsmath}
\date{\nonumber}
\begin{document}
\date{}
\newtheorem{theorem}{Theorem}
\newtheorem{proposition}{Proposition}
\newtheorem{lemma}{Lemma}
\newtheorem{definition}{Definition}

\renewcommand{\theequation}
{\arabic{section}.\arabic{equation}}

\renewcommand{\theequation}
{\arabic{section}.\arabic{equation}}
\title{A new extended matrix KP hierarchy and its solutions}
\maketitle
\begin{center}
{\it Yehui Huang$^{*}$\footnote{Corresponding author: Yehui Huang,
Tel: +86-13810446869, e-mail: huangyh@mails.tsinghua.edu.cn },
Xiaojun Liu$^{**}$\footnote{tigertooth4@gmail.com}, Yuqin
Yao$^{**}$\footnote{yyqinw@126.com} and Yunbo
Zeng$^{*}$\footnote{yzeng@math.tsinghua.edu.cn} }
\end{center}
\begin{center}{\small \it $^{*}$Department of Mathematical Science,
Tsinghua University, Beijing, 100084 , PR China\\
$^{**}$Department of
  Applied Mathematics, China Agricultural University, Beijing, 100083, PR China}
\end{center}
\begin{abstract}
With the square eigenfunctions symmetry constraint, we introduce a
new extended matrix KP hierarchy and its Lax representation from the
matrix KP hierarchy by adding a new $\tau_B$ flow. The extended KP
hierarchy contains two time series ${t_A}$ and ${\tau_B}$ and
eigenfunctions and adjoint eigenfunctions as components. The
extended matrix KP hierarchy and its $t_A$-reduction and $\tau_B$
reduction include two types of matrix KP hierarchy with
self-consistent sources and two types of (1+1)-dimensional reduced
matrix KP hierarchy with self-consistent sources. In particular, the
first type and second type of the 2+1 AKNS equation and the
Davey-Stewartson equation with self-consistent sources are deduced
from the extended matrix KP hierarchy. The generalized dressing
approach for solving the extended matrix KP hierarchy is proposed
and some solutions are presented. The soliton solutions of two types
of 2+1-dimensional AKNS equation with self-consistent sources and
two types of Davey-Stewartson equation with self-consistent sources
are studied.
\end{abstract}


\vskip .1cm

\section{Introduction}
\setcounter{equation}{0} Generalizations of Kadomtsev-Petviashvili
(KP) hierarchy attracts lots of interests from physical and
mathematical points of view $^{1-9}$. One kind of generalization is
the so called multi-component KP (mcKP) hierarchy or matrix KP
hierarchy $^{2-4}$, which contains many physical relevant nonlinear
integrable systems, such as (2+1)-dimensional AKNS hierarchy and
Davey-Stewartson (DS) equation. An extended DS equation can be
derived from matrix KP hierarchy $^{4}$. The explicit solutions of
the matrix KP equation were studied in $^{10-11}$. The relation
between the 1+1 dimensional C-integrable B\"{u}rgers hierarchy and
the matrix KP hierarchy is discussed in $^{12}$. The constrained
matrix KP flows was obtained from the matrix KP hierachy $^{3}$.
Another kind of generalization of KP equation is the so called KP
equation with self-consistent sources initiated by Mel'nikov
$^{5-7}$. The first type and second type of KP equation with
self-consistent sources were studied in $^{5}$. The KP equation with
self-consistent sources describes the interaction of a long wave
with a short-wave packet propagating on the $x$, $y$ plane at an
angle to each other. Recently, a systematic approach inspired by
squared eigenfunction symmetry constraint was proposed to construct
a new extended KP hierarchy $^{13}$. The extended KP hierarchy
extended the KP hierarchy by containing two times series $t_n$ and
$\tau_k$ and more components given by eigenfunctions and adjoint
eigenfunctions. This extended KP hierarchy and its $t_n$- and
$\tau_k$-reduction provide an unified way to find the two types of
KP equation with self-consistent sources and some (1+1)-dimensional
soliton equations with self-consistent sources. The solutions of the
extended KP hierarchy and extended mKP hierarchy can be derived
under a generalized dressing approach $^{14}$.

In this paper, we will construct the extension of the matrix KP
hierarchy. Inspired by the square eigenfunction symmetry constraint
of matrix KP hierarchy, we introduce a new $\tau_B$ flow by "extending" a
specific $t_A$-flow of matrix KP hierarchy. The extended matrix KP
hierarchy consists of $t_A$-flow, $\tau_B$-flow and the
$t_A$-evolutions of eigenfunctions and adjoint eigenfunctions. We
get the zero curvature representations for the extended matrix KP
hierarchy from the commutativity of $t_A$-flow and $\tau_B$-flow and its Lax representation.
The extended matrix KP hierarchy contains two time series ${t_A}$ and ${\tau_B}$ and more components by adding eigenfunctions and adjoint eigenfunctions,
and admits $t_A$-reduction and $\tau_B$-reduction. The extended matrix KP hierarchy and its two reduction provide an unified way to find two types of
matrix KP hierarchy with self-consistent sources, and two types of (1+1)-dimensional reduced matrix KP hierarchy with self-consistent sources.

By restricting the elements of the matrix KP hierarchy in $2\times
2$ matrices, we deduce two types of 2+1-dimensional AKNS equation
with self-consistent sources and two types of DS equation with
self-consistent sources from the extended matrix KP hierarchy as
examples. We can deduce two types of 1+1 AKNS equation with
self-consistent sources under two types of reductions. The DS
equation with self-consistent sources were also studied in $^{15}$
and $^{16}$. We would like to emphasize that our DS equation with
self-consistent sources is different from those in $^{15}$ and
$^{16}$ as they have different types of sources and the sources
satisfy different conditions.

The dressing method is an important tool for solving Gelfand Dickey
and KP hierarchy $^{17}$. However the dressing method for the matrix
KP hierarchy can not be directly applied to the extended matrix KP
hierarchy. With the combination of a dressing approach and the
method of variation of constants, we propose a generalized dressing
method for extended matrix KP hierarchy. By using this method, we
can solve the entire hierarchy of extended matrix KP hierarchy. We
solve two types of 2+1 AKNS equation with self-consistent sources
and two types of DS equation with self-consistent sources as
examples.

This paper is organized as follows. In section 2, we construct the
extended matrix KP hierarchy and derive its Lax pair, including two
types of the 2+1 AKNS equation with self-consistent sources and DS
equation with self-consistent sources as examples. In section 3,
$t_A$-reduction and $\tau_B$-reduction of the extended matrix KP
hierarchy are given. In section 4, a generalized dressing method for
the extended matrix KP hierarchy is discussed. In section 5, we give
the N-soliton solutions for the extended matrix KP hierarchy. The
soliton solution of 2+1 AKNS equation with self-consistent sources
and DS equation with self-consistent sources are studied. In section
6, we present the conclusion.

\section{The extended matrix KP hierarchy}
\setcounter{equation}{0} First we review the well known matrix KP
hierarchy $^{3-4}$. Let $g_0$ be a finite dimensional matrix
algebra, which means that the elements in $g_0$ are $N\times N$
matrices. The matrix KP hierarchy can be formulated by a
pseudodifferential operator which is called the dressing operator
$^{17}$
\begin{equation}
W=1+w_1\partial^{-1}+w_2\partial^{-2}+\cdots
\end{equation}
where $\partial=\frac{\partial}{\partial x}$, and the matrix valued
coefficients $w_i\in g_o$ are functions of $x$. $w=(w_1,w_2,\cdots)$
will be the dynamical fields of the matrix KP hierarchy. Define
$$\mathcal{A}=\{c_n\partial^n,\quad n\in \mathds{N},\quad c_n\in
g_0,\quad [c_n,~c_m]=0\}.$$ For each element $A\in\mathcal{A}$, the
evolution of $W$ is given by
\begin{equation}
W_{t_A}=-P_{<0}(WAW^{-1})W=-WA+M_AW,
\end{equation}
where
\begin{equation}
M_A=P_{\geq0}(WAW^{-1}),
\end{equation}
$P_{\geq0}(L)$ and $P_{<0}(L)$ denote
the nonnegative parts and negative parts of psedodifferential operator $L$. If we fix an arbitrary element
$C\in\mathcal{A}$ and define a Lax operator
\begin{equation}
L=WCW^{-1},
\end{equation}
the Lax equation of the matrix KP hierarchy is given by
\begin{equation}
L_{t_A}=[M_A,~L].
\end{equation}

The commutativity of $\partial_{t_A}$ and $\partial_{t_B}$ flows
gives rise to the zero-curvature equations of matrix KP hierarchy.
\begin{equation}
M_{A,t_B}-M_{B,t_A}+[M_A,~M_B]=0.
\end{equation}

It is known $^{3}$ that the evolution of $W$ given by
\begin{subequations}
\begin{eqnarray}
&&W_z=-\sum_{i=1}^{N}{\Phi_i\partial^{-1}\Psi_i^T}W,\\
&&\Phi_{i,t_A}=M_A(\Phi_i),\\
&&\Psi_{i,t_A}=-M_A^*(\Psi_i)
\end{eqnarray}
\end{subequations}
is compatible with the matrix KP hierarchy (2.5) and reduces the
matrix KP hierarchy to the constrained matrix KP hierarchy. If
$M_A=\sum_{i=0}^Nu_i\partial^i$, the adjoint operator $M_A^*$ is
defined by
\begin{equation}
M_A^*=\sum_{i=0}^N(-1)^i\partial^iu_i^T.
\end{equation}

Based on this observation, we now introduce a new $\tau_B$ flow given by
\begin{subequations}
\begin{eqnarray}
&&W_{\tau_B}=-P_{<0}(WBW^{-1})W+\sum_{i=1}^{N}{\Phi_i\partial^{-1}\Psi_i^T}W,\\
&&\Phi_{i,t_A}=M_A(\Phi_i),\\
&&\Psi_{i,t_A}=-M_A^*(\Psi_i).
\end{eqnarray}
\end{subequations}

We have the following lemma.

$\bf{Lemma~1.}$ $P_{<0}[M_A,\Phi\partial^{-1}\Psi^T]=M_A(\Phi)\partial^{-1}\Psi^T-\Phi\partial^{-1}(M_A^*(\Psi))^T.$

$\bf{Proof.}$ Without loss of generality, we consider a monomial
$Q=u\partial^k,~u\in g_0$. Using (2.8), we have
\begin{eqnarray}
P_{<0}(\partial^{-1}\Psi^TQ)&=&P_{<0}(\partial^{-1}\Psi^Tu\partial^k)\notag\\
&=&P_{<0}(\partial^{-1}\partial\Psi^Tu\partial^{k-1}-\partial^{-1}\partial(\Psi^Tu)\partial^{k-1})\notag\\
&=&-P_{<0}(\partial^{-1}\partial(\Psi_Tu)\partial^{k-1})=\cdots\notag\\
&=&(-1)^k\partial^{-1}(\Psi^Tu)^{(k)}=(-1)^k\partial^{-1}((u^T\Psi)^{(k)})^T\notag\\
&=&\partial^{-1}(Q^*(\Psi))^T.\notag
\end{eqnarray}
So we have
\begin{eqnarray}
P_{<0}[M_A,\Phi\partial^{-1}\Psi^T]&=&P_{<0}(M_A\Phi\partial^{-1}\Psi^T)-P_{<0}(\Phi\partial^{-1}\Psi^TM_A)\notag\\
&=&M_A(\Phi)\partial^{-1}\Psi^T-\Phi\partial^{-1}(M_A^*(\Psi))^T.\notag\quad\quad\Box
\end{eqnarray}

Further more, we find that
\begin{eqnarray}
L_{\tau_B}&=&W_{\tau_B}CW^{-1}-WCW^{-1}W_{\tau_B}W^{-1}\notag\\
&=&M_BL-LM_B+\Phi\partial^{-1}\Psi^TL-L\Phi\partial^{-1}\Psi^T\\
&=&[M_B+\Phi\partial^{-1}\Psi^T,~L].\notag
\end{eqnarray}

$\bf{Lemma~2.}$ The $\tau_B$ flow given by (2.9) and (2.10) is
compatible with the matrix KP hierarchy (2.5), namely,
$(W_{t_A})_{\tau_B}=(W_{\tau_B})_{t_A}$,
$(L_{t_A})_{\tau_B}=(L_{\tau_B})_{t_A}$.

$\bf{Proof.}$ For convenience, we omit $\sum$. Notice that $W_{\tau_B}=W_{t_B}+\Phi\partial^{-1}\Psi^TW$, we obtain
\begin{eqnarray}
(W_{\tau_B})_{t_A}&=&(W_{t_B})_{t_A}+(M_A(\Phi)\partial^{-1}\Psi^T-\Phi\partial^{-1}(M_A^*(\Psi))^T-\Phi\partial^{-1}\Psi^TP_{<0}(WAW^{-1}))W,\notag\\
(W_{t_A})_{\tau_B}&=&-P_{<0}(W_{t_B}AW^{-1}+\Phi\partial^{-1}\Psi^TWAW^{-1}-WAW^{-1}W_{t_B}W^{-1}\notag\\
&&-WAW^{-1}\Phi\partial^{-1}\Psi^T)W-P_{<0}(WAW^{-1})W_{t_B}-P_{<0}(WAW^{-1})\Phi\partial^{-1}\Psi^TW,\notag
\end{eqnarray}
so
\begin{eqnarray}
(W_{\tau_B})_{t_A}-(W_{t_A})_{\tau_B}&=&(W_{t_B})_{t_A}-(W_{t_A})_{t_B}+(M_A(\Phi)\partial^{-1}\Psi^T-\Phi\partial^{-1}(M_A^*(\Psi))^T\notag\\
&&-[\Phi\partial^{-1}\Psi^T,~P_{<0}(WAW^{-1})]+P_{<0}([\Phi\partial^{-1}\Psi^T,~WAW^{-1}]))W.\notag
\end{eqnarray}
Using Lemma 1, we find that
\begin{eqnarray}
P_{<0}([\Phi\partial^{-1}\Psi^T,~WAW^{-1}])&=&P_{<0}(\Phi\partial^{-1}\Psi^TM_A+\Phi\partial^{-1}\Psi^TP_{<0}(WAW^{-1})\notag\\
&-&M_A\Phi\partial^{-1}\Psi^T-P_{<0}(WAW^{-1})\Phi\partial^{-1}\Psi^T)\notag\\
&=&[\Phi\partial^{-1}\Psi^T,~P_{<0}(WAW^{-1})]+\Phi\partial^{-1}(M_A^*(\Psi))^T-M_A(\Phi)\partial^{-1}\Psi^T,\notag
\end{eqnarray}
so we have that $(W_{\tau_B})_{t_A}-(W_{t_A})_{\tau_B}=0$. As
$(L_{t_A})_{\tau_B}=((WCW^{-1})_{t_A})_{\tau_B}$,
$(L_{\tau_B})_{t_A}=((WCW^{-1})_{\tau_B})_{t_A}$, we find
$(L_{t_A})_{\tau_B}=(L_{\tau_B})_{t_A}$ from
$(W_{t_A})_{\tau_B}=(W_{\tau_B})_{t_A}$. $\Box$

The commutativity of $t_A$ flow and $\tau_B$ flow enables us to obtain a new extended matrix KP hierarchy as
\begin{subequations}
\begin{eqnarray}
&&\partial_{t_A}L=[M_A,~L],\\
&&\partial_{\tau_B}L=[M_B+\sum_{i=1}^{N}{\Phi_i\partial^{-1}\Psi_i^T},~L]\\
&&\Phi_{i,t_A}=M_A(\Phi_i),~i=1,\ldots,N,\\
&&\Psi_{i,t_A}=-M_A^*(\Psi_i),~i=1,\ldots,N.
\end{eqnarray}
\end{subequations}

$\bf{Proposition~1}.$ The commutativity of (2.11a) and (2.11b) under
(2.11c) and (2.11d) gives rise to the zero-curvature equation for
extended matrix KP hierarchy (2.11)
\begin{subequations}
\begin{eqnarray}
&&M_{A,\tau_B}-(M_B+\sum_{i=1}^N\Phi_i\partial^{-1}\Psi_i^T)_{t_A}+[M_A,~M_B+\sum_{i=1}^{N}{\Phi_i\partial^{-1}\Psi_i^T}]=0,\\
&&\Phi_{i,t_A}=M_A(\Phi_i),\\
&&\Psi_{i,t_A}=-M_A^*(\Psi_i),~i=1,\ldots,N,
\end{eqnarray}
\end{subequations}
or equivalently
\begin{subequations}
\begin{eqnarray}
&&M_{A,\tau_B}-M_{B,t_A}+[M_A,~M_B]-\sum_{i=1}^{N}{P_{\geq0}}([\Phi_i\partial^{-1}\Psi_i^T,~M_A])=0,\\
&&\Phi_{i,t_A}=M_A(\Phi_i),\\
&&\Psi_{i,t_A}=-M_A^*(\Psi_i),~i=1,\ldots,N,
\end{eqnarray}
\end{subequations}
with the Lax representation given by
\begin{subequations}
\begin{eqnarray}
&&\psi_{t_A}=M_A(\psi),\\
&&\psi_{\tau_B}=(M_B+\sum_{i=1}^{N}{\Phi_i\partial^{-1}\Psi_i^T})(\psi).
\end{eqnarray}
\end{subequations}

$\bf{Proof.}$ The commutativity of (2.11a) and (2.11b) under (2.11c)
and (2.11d) immediately gives rise to (2.12). Now we prove (2.13).
Using Lemma 1, we have
\begin{eqnarray}
-(\Phi\partial^{-1}\Psi^T)_{t_A}+[M_A,~\Phi\partial^{-1}\Psi^T]&=&-M_A(\Phi)\partial^{-1}\Psi^T+\Phi\partial^{-1}M_A^*(\Psi)^T+[M_A,~\Phi\partial^{-1}\Psi^T]\notag\\
&=&-P_{<0}[M_A,~\Phi\partial^{-1}\Psi^T]+[M_A,~\Phi\partial^{-1}\Psi^T]\\
&=&P_{\geq0}[M_A,~\Phi\partial^{-1}\Psi^T].~~~~\Box\notag
\end{eqnarray}

$\bf{Remark.}$ The extended matrix KP hierarchy extends the matrix KP hierarchy by containing two time series ${t_A}$ and ${\tau_B}$
and more components $\Phi_i$ and $\Psi_i$, $i=1,\ldots,N$.

In the following we restrict $w_i$ to $2\times 2$ matrices and
consider the extended matrix KP hierarchy with $A=\sigma_3\partial$,
$B=\sigma_3\partial^2$ and $C=\sigma_3\partial$, where
$\sigma_3=\left(\begin{array}{cc}1 & 0\\0 & -1\\\end{array}\right)$.
The associated times are $t_A=y$, $t_B=t$. We introduce
$U=\left(\begin{array}{cc}0 & r\\q &
0\\\end{array}\right)=[\omega_1,~\sigma_3]=-2\sigma_3\omega_1^{off},\quad
D=\left(\begin{array}{cc}a & 0\\0 &
b\\\end{array}\right)=-2\omega_1^{diag}.$ The zero-curvature
equation (2.6) leads to the evolution equation
\begin{subequations}
\begin{eqnarray}
&&U_t=\frac{1}{2}\sigma_3(U_{xx}+U_{yy})+\sigma_3U^3+[D_y,U],\\
&&\sigma_3D_y-D_x+U^2=0,
\end{eqnarray}
\end{subequations}
i.e.
\begin{subequations}
\begin{eqnarray}
&&r_t=\frac{1}{2}(r_{xx}+r_{yy})+r^2q+(a_y-b_y)r,\quad a_y-a_x+rq=0,\\
&&q_t=-\frac{1}{2}(q_{xx}+q_{yy})-q^2r-(a_y-b_y)q,\quad b_y+b_x-rq=0,
\end{eqnarray}
\end{subequations}
As $a_x-a_y=b_y+b_x$, we may assume $a+b=\phi_x$, $a-b=\phi_y$,
which leads to $\phi_{xx}-\phi_{yy}=2rq$. Denote $v=\phi_y$, then we
have
\begin{subequations}
\begin{eqnarray}
&&r_t=\frac{1}{2}(r_{xx}+r_{yy})+r^2q+v_yr,\\
&&q_t=-\frac{1}{2}(q_{xx}+q_{yy})-q^2r-v_yq,\\
&&v_{xx}-v_{yy}=2(rq)_y,
\end{eqnarray}
\end{subequations}
which is $2+1$ dimensional AKNS equation $^{3}$. If we replace $U_t$
by $iU_t$ and assume that $q=\bar{r}$, $\phi=\bar{\phi}$, then the
above system reduces to the Davey-Stewartson I (DSI) equation
\begin{equation}
ir_t=\frac{1}{2}(r_{xx}+r_{yy})+|r|^2r+v_yr,\quad
v_{xx}-v_{yy}=2(|r|^2)_y.
\end{equation}

Now we derive the (2+1) dimensional AKNS equation with
self-consistent sources and Davey-Stewartson equation with
self-consistent sources from (2.13).

$\bf{Example~1.}$ If we take $t_A=y$ and $\tau_B=t$, we obtain the
first type of $2+1$ dimensional AKNS equation with self-consistent
sources from (2.13)
\begin{subequations}
\begin{eqnarray}
&&U_t=\frac{1}{2}\sigma_3(U_{xx}+U_{yy})+\sigma_3U^3+[D_y,U]+[\sum_{i=1}^N{\Phi_i\Psi_i^T},\sigma_3],\\
&&\sigma_3D_y-D_x+U^2=0,\\
&&\Phi_{i,y}=\sigma_3\Phi_{i,x}+U\Phi_i,\\
&&\Psi_{i,y}=\sigma_3\Psi_{i,x}-U^T\Psi_i,
\end{eqnarray}
\end{subequations}
or we can rewrite it as
\begin{subequations}
\begin{eqnarray}
&&r_t=\frac{1}{2}(r_{xx}+r_{yy})+r^2q+v_yr-\sum_{i=1}^N(2(\phi_{i11}\psi_{i21}+\phi_{i12}\psi_{i22})),\\
&&q_t=-\frac{1}{2}(q_{xx}+q_{yy})-q^2r-v_yq+\sum_{i=1}^N(2(\phi_{i21}\psi_{i11}+\phi_{i22}\psi_{i12})),\\
&&v_{xx}-v_{yy}=2(rq)_y,\\
&&\Phi_{i,y}=\sigma_3\Phi_{i,x}+U\Phi_i,\\
&&\Psi_{i,y}=\sigma_3\Psi_{i,x}-U^T\Psi_i.
\end{eqnarray}
\end{subequations}
Here and afterward $\Phi_i=\left(\begin{array}{cc}\phi_{i11} & \phi_{i12} \\ \phi_{i21} & \phi_{i22} \\\end{array}\right)$,
$\Psi_i=\left(\begin{array}{cc}\psi_{i11} & \psi_{i12} \\ \psi_{i21} & \psi_{i22} \\\end{array}\right)$, $i=1,\ldots,N$ are $2\times2$ matrices.

Its Lax representation is
\begin{subequations}
\begin{eqnarray}
&&\psi_y=(\sigma_3\partial+U)\psi,\\
&&\psi_t=(\sigma_3\partial^2+U\partial+\frac{1}{2}U_x+\frac{1}{2}\sigma_3U^2+\frac{1}{2}\sigma_3U_y+D_y+\sum_{i=1}^N\Phi_i\partial^{-1}\Psi_i^T)\psi.
\end{eqnarray}
\end{subequations}

$\bf{Example~2.}$ When we take $\tau_A=y$ and $t_B=t$, we get the
second type of $2+1$ dimensional AKNS equation with self-consistent
sources from (2.13)
\begin{subequations}
\begin{eqnarray}
&&U_t=\frac{1}{2}\sigma_3(U_{xx}+U_{yy})+\sigma_3U^3+[D_y,U]-\sum_{i=1}^N\left([U,(\Phi_i\Psi_i^T)^{diag}]+2\sigma_3(\Phi_{i,x}\Psi_i^T)^{off}\right),\\
&&(\sigma_3D_y-D_x+U^2)_y-\sum_{i=1}^N\left([U,(\Phi_i\Psi_i^T)^{off}]+2\sigma_3(\Phi_i\Psi_i^T)_x^{diag}\right)=0,\\
&&\Phi_{i,t}=\sigma_3\Phi_{i,xx}+U\Phi_{i,x}+\frac{1}{2}(U_x+\sigma_3U^2+\sigma_3U_y+D_y)\Phi_i,~i=1,\ldots,N,\\
&&\Psi_{i,t}=-\sigma_3\Psi_{i,xx}+U^T\Psi_{i,x}-\frac{1}{2}(U_x+\sigma_3U^2+\sigma_3U_y+D_y)^T\Psi_i,~i=1,\ldots,N,
\end{eqnarray}
\end{subequations}
or
\begin{subequations}
\begin{eqnarray}
r_t&=&\frac{1}{2}(r_{xx}+r_{yy})+r^2q+v_yr+\sum_{i=1}^N(r(\phi_{i11}\psi_{i11}+\phi_{i12}\psi_{i12}-\phi_{i21}\psi_{i21}-\phi_{i22}\psi_{i22})\notag\\
&&-2(\phi_{i11,x}\psi_{i21}+\phi_{i12,x}\psi_{i22})),\\
q_t&=&-\frac{1}{2}(q_{xx}+q_{yy})-q^2r-v_yq-\sum_{i=1}^N(q(\phi_{i11}\psi_{i11}+\phi_{i12}\psi_{i12}-\phi_{i21}\psi_{i21}-\phi_{i22}\psi_{i22})\notag\\
&&-2(\phi_{i21,x}\psi_{i11}+\phi_{i22,x}\psi_{i12})),\\
v_{yy}-v_{xx}&=&2(\sum_{i=1}^N(r(\phi_{i21}\psi_{i11}+\phi_{i22}\psi_{i12}-q(\phi_{i11}\psi_{i21}+\phi_{i12}\psi_{i22})\notag\\
&&+2(\phi_{i11}\psi_{i11}+\phi_{i12}\psi_{i12})_x))),\\
\Phi_{i,t}&=&\sigma_3\Phi_{i,xx}+U\Phi_{i,x}+\frac{1}{2}(U_x+\sigma_3U^2+\sigma_3U_y+D_y)\Phi_i,~i=1,\ldots,N,\\
\Psi_{i,t}&=&-\sigma_3\Psi_{i,xx}+U^T\Psi_{i,x}-\frac{1}{2}(U_x+\sigma_3U^2+\sigma_3U_y+D_y)^T\Psi_i,~i=1,\ldots,N,
\end{eqnarray}
\end{subequations}

Its Lax representation is
\begin{subequations}
\begin{eqnarray}
&&\psi_y=(\sigma_3\partial+U+\sum_{i=1}^N\Phi_i\partial^{-1}\Psi_i^T)\psi,\\
&&\psi_t=(\sigma_3\partial^2+U\partial+\frac{1}{2}U_x+\frac{1}{2}\sigma_3U^2+\frac{1}{2}\sigma_3U_y+D_y)\psi.
\end{eqnarray}
\end{subequations}

$\bf{Example~3.}$ Define $\tilde{U}=\left(\begin{array}{cc}0 & r \\
\bar{r} & 0 \\\end{array}\right)$, the first type of DSI equation
with self-consistent sources is
\begin{subequations}
\begin{eqnarray}
&&ir_t=\frac{1}{2}(r_{xx}+r_{yy})+|r|^2r+v_yr-2\sum_{j=1}^N(\phi_{j11}\psi_{j21}+\phi_{j12}\psi_{j22}),\\
&&v_{xx}-v_{yy}=2(|r|^2)_y,\\
&&\Phi_{j,y}=\sigma_3\Phi_{j,x}+\tilde{U}\Phi_j,~j=1,\ldots,N,\\
&&\Psi_{j,y}=\sigma_3\Psi_{j,x}-\tilde{U}^T\Psi_j,~j=1,\ldots,N.
\end{eqnarray}
\end{subequations}

Its Lax representaion is
\begin{subequations}
\begin{eqnarray}
&&\psi_y=(\sigma_3\partial+\tilde{U})\psi,\\
&&\psi_t=-i(\sigma_3\partial^2+\tilde{U}\partial+\frac{1}{2}\tilde{U}_x
+\frac{1}{2}\sigma_3\tilde{U}^2+\frac{1}{2}\sigma_3\tilde{U}_y+D_y+\sum_{j=1}^N\Phi_j\partial^{-1}\Psi_j^T)\psi.
\end{eqnarray}
\end{subequations}

$\bf{Example~4.}$ The second type of DSI equation with self-consistent sources is
\begin{subequations}
\begin{eqnarray}
ir_t&=&\frac{1}{2}(r_{xx}+r_{yy})+|r|^2r+v_yr+\sum_{j=1}^N(r(\phi_{j11}\psi_{j11}+\phi_{j12}\psi_{j12}-\phi_{j21}\psi_{j21}-\phi_{j22}\psi_{j22})\notag\\
&&-2(\phi_{j11,x}\psi_{j21}+\phi_{j12,x}\psi_{j22})),\\
v_{yy}-v_{xx}&=&2(\sum_{j=1}^N(r(\phi_{j21}\psi_{j11}+\phi_{j22}\psi_{j12}-\bar{r}(\phi_{j11}\psi_{j21}+\phi_{j12}\psi_{j22})\notag\\
&&+2(\phi_{j11}\psi_{j11}+\phi_{j12}\psi_{j12})_x))),\\
\Phi_{j,t}&=&\sigma_3\Phi_{j,xx}+\tilde{U}\Phi_{j,x}+\frac{1}{2}(\tilde{U}_x+\sigma_3\tilde{U}^2+\sigma_3\tilde{U}_y+D_y)\Phi_j,~j=1,\ldots,N,\\
\Psi_{j,t}&=&-\sigma_3\Psi_{j,xx}+\tilde{U}^T\Psi_{j,x}-\frac{1}{2}(\tilde{U}_x+\sigma_3\tilde{U}^2+\sigma_3\tilde{U}_y+D_y)^T\Psi_j,~j=1,\ldots,N.
\end{eqnarray}
\end{subequations}

Its Lax representation is
\begin{subequations}
\begin{eqnarray}
&&\psi_y=(\sigma_3\partial+\tilde{U}+\sum_{j=1}^N\Phi_j\partial^{-1}\Psi_j^T)\psi,\\
&&\psi_t=-i(\sigma_3\partial^2+\tilde{U}\partial+\frac{1}{2}\tilde{U}_x+\frac{1}{2}\sigma_3\tilde{U}^2+\frac{1}{2}\sigma_3\tilde{U}_y+D_y)\psi.
\end{eqnarray}
\end{subequations}

$\bf{Remark.}$ The extended matrix KP hierarchy (2.11) provides a
unified way to construct the first type and second type of the
(2+1)-dimensional AKNS equation (and DSI equation) with
self-consistent sources and their Lax representation.

\section{Reductions of the extended matrix KP hierarchy}
\setcounter{equation}{0} The extended KP hierarchy depends on two time series $t_A$ and $\tau_B$. It is natural to consider its $t_A$-reduction and $\tau_B$-reduction.
\subsection{The $t_A$-reduction}
The $t_A$-reduction of the extended matrix
KP hierarchy is given by
\begin{equation}
WAW^{-1}=M_A,
\end{equation}
where $A=C_k\partial^k$. The wave function and the adjoint wave function are given by
\begin{equation}
\Phi(t,z)=W\exp{(\xi(t,z))},~\Phi^*(t,z)=(W^*)^{-1}\exp{(-\xi(t,z))},
\end{equation}
where $\xi(t,z)=\sum_{i>0}t_iz^i$, $t_1=x$.
Then we have
\begin{eqnarray}
&&M_A(\Phi)=WAW^{-1}\Phi=z^k\Phi C_k,\\
&&M_A^*(\Phi^*)=(WAW^{-1})^*\Phi^*=-z^k\Phi^*C_k.\\
&&L_{t_A}=[M_A,~L]=[WAW^{-1},~WCW^{-1}]=W[A,~C]W^{-1}=0.
\end{eqnarray}
So $L$ is independent of $t_A$ and we can drop $t_A$ dependence from
(2.12)
\begin{subequations}
\begin{eqnarray}
&&(M_A)_{\tau_B}=[M_B+\sum_{i=1}^N\Phi_i\partial^{-1}\Psi_i^T,~M_A],\\
&&M_A(\Phi_i)=\lambda_i^k\Phi_iC_k,\\
&&M_A^*(\Psi_i)=\lambda_i^k\Psi_iC_k.
\end{eqnarray}
\end{subequations}

When $A=\sigma_3\partial$ and $B=\sigma_3\partial^2$ we have
$M_A(\Phi_i)=WAW^{-1}(\Phi_i)=\lambda_i\Phi_i\sigma_3.$
Then the first type of the $2+1$-dimensional AKNS equation with
self-consistent sources reduces to the first type of
$1+1$-dimensional AKNS equation with self-consistent sources
\begin{subequations}
\begin{eqnarray}
&&U_t=\frac{1}{2}\sigma_3U_{xx}+\sigma_3U^3+[\sum_{i=1}^N\Phi_i\Psi_i^T,~\sigma_3],\\
&&U^2=D_x,\\
&&\sigma_3\Phi_{i,x}+U\Phi_i=\lambda_i\Phi_i\sigma_3,\\
&&\sigma_3\Psi_{i,x}-U^T\Psi_i=\lambda_i\Psi_i\sigma_3.
\end{eqnarray}
\end{subequations}

\subsection{The $\tau_B$-reduction}
The $\tau_B$-reduction is given by $^{3}$
\begin{equation}
WBW^{-1}=M_B+\sum_{i=1}^N\Phi_i\partial^{-1}\Psi_i^T.
\end{equation}
Then we can drop $\tau_B$ dependence from (2.12)
\begin{subequations}
\begin{eqnarray}
&&(M_B+\sum_{i=1}^N\Phi_i\partial^{-1}\Psi_i^T)_{t_A}=[M_A,~M_B+\sum_{i=1}^N\Phi_i\partial^{-1}\Psi_i^T],\\
&&\Phi_{i,t_A}=M_A(\Phi_i),\\
&&\Psi_{i,t_A}=-M_A^*(\Psi_i).
\end{eqnarray}
\end{subequations}
which is just the constrained matrix KP hierarchy given in $^{3}$.

When $B=\sigma_3\partial$ and $A=\sigma_3\partial^2$, we get the constrained (2+1)-dimensional AKNS equation or second type of AKNS equation with self-consistent sources.
\begin{subequations}
\begin{eqnarray}
&&U_t=\frac{1}{2}\sigma_3U_{xx}+\sigma_3U^3+[\sum_{j=1}^m(\Phi_j\Psi_j^T)^{diag},~U]+\sigma_3\sum_{j=1}^m(\Phi_{jx}\Psi_j^T-\Phi_j\Psi_{jx}^T)^{off},\\
&&\Phi_{it}=\sigma_3\Phi_{ixx}+U\Phi_{ix}+\frac{1}{2}U_x\Phi_i+\frac{1}{2}\sigma_3U^2\Phi_i+\sum_{j=1}^m\Phi_j\Psi_j^T\Phi_i+\sum_{j=1}^m(\Phi_j\Psi_j^T)^{diag}\Phi_i,\\
&&\Psi_{it}=-\sigma_3\Psi_{ixx}+U^T\Psi_{ix}+\frac{1}{2}U_x^T\Psi_i-\frac{1}{2}\sigma_3U^2\Psi_i-\sum_{j=1}^m\Psi_j\Phi_j^T\Psi_i-\sum_{j=1}^m(\Psi_j\Phi_j^T)^{diag}\Psi_i.
\end{eqnarray}
\end{subequations}

$\bf{Remark.}$ The extended matrix KP hierarchy and its $t_A$-reduction and $\tau_B$-reduction provide a simple and unified way
to obtain the two types of (2+1)-dimensional and (1+1)-dimensional AKNS equation with self-consistent sources.

\section{Generalized dressing approach for extended matrix KP hierarchy}
\setcounter{equation}{0} In the following, we restrict $g_0$ to be
the matrix algebra of dimensional $2\times 2$. Now we propose a
generalized dressing approach for the extended matrix KP hierarchy.
For the dressing form of $L$ given by (2.4)
\begin{equation}
L=WCW^{-1},
\end{equation}
usually the $W$ has finite terms, so it is
equivalent to assume that the dressing operator $W$ is a pure
differential operator of order $N$ as follows
\begin{equation}
W=\partial^N+w_1\partial^{N-1}+w_2\partial^{N-2}+\cdots+w_N.
\end{equation}
Let $2\times 2$ matrices $f_i$, $g_i$ satisfy
\begin{equation}
f_{i,t_A}=A(f_i),~f_{i,\tau_B}=B(f_i),~g_{i,t_A}=A(g_i),~g_{i,\tau_B}=B(g_i),~i=1,\cdots,N.
\end{equation}
By means of the method of variation of constants, let $2\times 2$ matrices $h_i$ be the linear combination of $f_i$ and $g_i$ as
\begin{equation}
h_i=f_i+\alpha_i(\tau_B)g_i,\quad i=1,\cdots,N
\end{equation}
with $\alpha_i$ being a function of $\tau_B$. We assume that $h_i$
and its derivatives are invertible matrices and the $2N\times2N$
matrix $\left(\begin{array}{cccc}h_1 & h_2 & \cdots & h_N \\
h_1^{(1)} & h_2^{(1)} & \cdots & h_N^{(1)} \\ \vdots & \vdots &
\ddots & \vdots \\ h_1^{(N-1)} & h_2^{(N-1)} & \cdots & h_N^{(N-1)}
\\\end{array}\right)$ is invertible.

$\bf{Theorem~1}$ $^{3}$: Let $a_0,\ldots,a_{N-1}$ be the $2\times 2$
matrix functions determined as the solution of the linear algebraic
system
\begin{equation}
\sum_{i=0}^Na_ih_j^{(i)}=0,\quad j=1,\ldots,N,
\end{equation}
with $a_N=1$. Then $W=\sum_{i=0}^Na_i\partial^i$ satisfies (2.2) and
$L=WCW^{-1}$ satisfies the matrix KP hierarchy (2.5).

We have
\begin{equation}
W(h_i)=0,~i=1,\cdots,N
\end{equation}

$\bf{Theorem~2}$: Let $b_1,\ldots,b_N$ be the $2\times 2$ matrix functions satisfy
\begin{equation}
\sum_{j=1}^Nh_j^{(i)}b_j=\delta_{i,N-1},\quad i=0,\ldots,N-1.
\end{equation}
Define $\Phi_i=-\dot{\alpha}_iW(g_i)$, and $\Psi_i^T=b_i$, where
$\dot{\alpha}_i=d\alpha_i/d\tau_B$, then
$W=\sum_{i=0}^Na_i\partial^i$, $\Phi_i$, $\Psi_i$ and $L=WCW^{-1}$
satisfy extended matrix KP hierarchy (2.11).

To proof Theorem 2, we need several lemmas under the above
assumptions.

$\bf{Lemma~3}$: $W^{-1}=\sum_{i=1}^N{h_i\partial^{-1}\Psi_i^T}$.

Proof: Using(4.6) and (4.7), we have
\begin{eqnarray}
P_{\geq0}(W\sum_{i=1}^N{h_i\partial^{-1}\Psi_i^T})&=&P_{\geq0}(W\sum_{i=1}^N{h_i\partial^{-1}b_i})=P_{\geq0}(W\sum_{k=0}^{\infty}{\partial^{-k-1}\sum_{i=1}^N{h_i^{(k)}b_i}})\notag\\
&=&P_{\geq0}(W\partial^{-N}+\sum_{k=N}^{\infty}\partial^{-k-1}\sum_{i=1}^Nh_i^{(k)}b_i)\notag\\
&=&P_{\geq0}(W\partial^{-N}(1+\sum_{k=0}^{\infty}{\partial^{-k-1}\sum_{i=1}^N{h_i^{(N+k)}b_i}}))=1.\notag\\
P_{\leq0}(W\sum_{i=1}^N{h_i\partial^{-1}\Psi_i^T})&=&\sum_{i=1}^N(W(h_i))\partial^{-1}\Psi_i^T=0.
\end{eqnarray}
So we know that $W^{-1}=\sum_{i=1}^N{h_i\partial^{-1}\Psi_i^T}$.

$\bf{Lemma~4}$: $W^{*}(\Psi_i)=0$.

Proof: From the relation $W^{*}(W^{-1})^{*}\partial^j=\partial^j$,
we know that
$$0=\textrm{Res}_{\partial}W^{*}(\sum_{i=1}^N{h_i\partial^{-1}\Psi_i^T})^{*}\partial^j=-\textrm{Res}_{\partial}W^{*}\sum_{i=1}^N{\Psi_i\partial^{-1}h_i^T\partial^j}=(-1)^{j+1}\sum_{i=1}^N{W^{*}(\Psi_i)h_i^{T(j)}}.$$
As the $2N\times2N$
matrix $\left(\begin{array}{cccc}h_1 & h_2 & \cdots & h_N \\
h_1^{(1)} & h_2^{(1)} & \cdots & h_N^{(1)} \\ \vdots & \vdots &
\ddots & \vdots \\ h_1^{(N-1)} & h_2^{(N-1)} & \cdots & h_N^{(N-1)}
\\\end{array}\right)$ is invertible, we find that $W^{*}(\Psi_i)=0$.

$\bf{Lemma~5}$: The operator $\partial^{-1}\Psi_i^TW$ is a pure
differential operator for each $i$. Further more, for $1\leq i,
j\leq N$, $(\partial^{-1}\Psi_i^TW)(h_j)=\delta_{ij}I$.

Proof: As
$P_{<0}(\partial^{-1}\Psi_i^TW)=\partial^{-1}(W^{*}(\Psi_i))^T=0$, we
know that $\partial^{-1}\Psi_i^TW$ is a pure differential operator.
Let $c_{ij}=(\partial^{-1}\Psi_i^TW)(h_j)$. We find that
$\partial(c_{ij})=\Psi_i^TW(h_j)=0$ and
$$\sum_{i=1}^Nh_i^{(k)}c_{ij}=\partial^k(\sum_ih_ic_{ij})=\partial^k(\sum_i(h_i\partial^{-1}\Psi_i^TW)(h_j))=h_j^{(k)}.$$
So we have $c_{ij}=\delta_{ij}I$.

$\bf{Proposition~2}$: $W$ satisfies
$W_{\tau_B}=-P_{<0}(WBW^{-1})W+\sum_{i=1}^N{\Phi_i\partial^{-1}\Psi_i^TW}$.

Proof: Taking $\partial_{\tau_B}$ to the identity $W(h_i)=0$ and
using (4.3), Lemma 3 and Lemma 5, we have
\begin{eqnarray}
0&=&W_{\tau_B}(h_i)+WB(h_i)+\dot{\alpha}_iW(g_i)\notag\\
&=&(\partial_{\tau_B}W)(h_i)+(WBW^{-1}W)(h_i)-\sum_{j=1}^N{\Phi_j\delta_{ji}}\notag\\
&=&(\partial_{\tau_B}W+P_{<0}(WBW^{-1})W-\sum_{j=1}^N{\Phi_j\partial^{-1}\Psi_j^TW})(h_i).
\end{eqnarray}
Since the non-negative operator acting on $h_i$ above has degree lower than $N$ and $h_i$ are $N$ independent functions, the operator itself must be zero.
Hence the proposition is proven.

Proof of the Theorem 2: The proof of (2.11a) is similar as we have
in Theorem 1. By using (2.10), we get (2.11b). By a direct
calculation, we get (2.11c) and (2.11d).

\section{Solutions for extended matrix KP hierarchy}
\setcounter{equation}{0} By using the generalized dressing approach
given by Theorem 2, we can construct the
explicit solutions to the extended matrix KP hierarchy.

For the (2+1)-dimensional AKNS equation with self-consistent
sources, we choose proper $2\times2$ matrices $f_i$ and $g_i$. The
solution of $f_y=\sigma_3f_x,~f_t=\sigma_3f_{xx}$ is
\begin{equation}
f=\left(\begin{array}{cc}\frac{c_{11}}{\lambda}+d_{11}e^{\lambda
x+\lambda y+\lambda^2t}
& \frac{c_{12}}{\mu}+d_{12}e^{\mu x+\mu y+\mu^2t}\\
\frac{c_{21}}{\lambda}+d_{21}e^{\lambda x-\lambda y-\lambda^2t} &
\frac{c_{22}}{\mu}+d_{22}e^{\mu x-\mu y-\mu^2t}\\\end{array}\right),\\
\end{equation}
where $c_{ij},~d_{ij}$ are arbitrary constants, but they should be
chosen such that $f$ is invertible for any $x,y,t$. We can define
\begin{eqnarray}
&&f_i=\left(\begin{array}{cc}\frac{1}{\lambda_i}+e^{\lambda_ix+\lambda_iy+\lambda_i^2t}
& 0\\0 &
\frac{1}{\mu_i}+e^{\mu_ix-\mu_iy-\mu_i^2t}\\\end{array}\right),\\
&&g_i=\left(\begin{array}{cc}0 &
-e^{\mu_ix+\mu_iy+\mu_i^2t}\\e^{\lambda_ix-\lambda_iy-\lambda_i^2t}
& 0\\\end{array}\right).
\end{eqnarray}
where $\lambda_i>0$, $\mu_i>0$. In this way we have $a_i$ and $b_i$
and the N-soliton solution of the (2+1)-dimensional AKNS equation
with self-consistent sources is
\begin{equation}
U=-2\sigma_3a_{n-1}^{off},~D=a_{n-1}^{diag},~\Phi_i=-\dot{\alpha}_iW(g_i),~\Psi_i=b_i,~i=1,\ldots,n.
\end{equation}

The one-soliton solution of the first type of $(2+1)$-AKNS equation
with a self-consitent source (2.21) can be constructed from
$h=f+\alpha(t) g=\left(\begin{array}{cc}\frac{1}{\lambda}+e^{\lambda
x+\lambda y+\lambda^2t} &
-\alpha(t)e^{\mu x+\mu y+\mu^2t}\\
\alpha(t)e^{\lambda x-\lambda y-\lambda^2t} & \frac{1}{\mu}+e^{\mu
x-\mu
y-\mu^2t}\\\end{array}\right):=\left(\begin{array}{cc}\frac{1}{\lambda}+e^{\xi_{11}}
& -\alpha(t)e^{\xi_{12}}\\\alpha(t)e^{\xi_{21}} &
\frac{1}{\mu}+e^{\xi_{22}}\\\end{array}\right)$¸ø³ö¡£

From Theorem 1 and Theorem 2 we have
$a_0=-h_xh^{-1},~a_1=1,~b_0=h^{-1},$ which gives rise to
\begin{eqnarray}
&&W=\partial-h_xh^{-1},~U=2\sigma_3(h_xh^{-1})^{off},~D=2(h_xh^{-1})^{diag},\notag\\
&&\Phi=-\dot{\alpha}(t)(g_x-h_xh^{-1}g),~\Psi^T=h^{-1}.
\end{eqnarray}

So the one-soliton of the first type of (2+1)-AKNS equation with a
self-consistent source (2.21) is
\begin{align}
&q=-2\alpha(t)\frac{\frac{\lambda}{\mu}e^{\xi_{21}}+(\lambda-\mu)e^{\xi_{21}+\xi_{22}}}
{(\frac{1}{\lambda}+e^{\xi_{11}})(\frac{1}{\mu}+e^{\xi_{22}})+\alpha(t)^2e^{\xi_{12}+\xi_{21}}},\\
&r=2\alpha(t)\frac{-\frac{\mu}{\lambda}e^{\xi_{12}}+(\lambda-\mu)e^{\xi_{11}+\xi_{12}}}
{(\frac{1}{\lambda}+e^{\xi_{11}})(\frac{1}{\mu}+e^{\xi_{22}})+\alpha(t)^2e^{\xi_{12}+\xi_{21}}},\\
&v=2\frac{\frac{\lambda}{\mu}e^{\xi_{11}}-\frac{\mu}{\lambda}e^{\xi_{22}}+(\lambda-\mu)(e^{\xi_{11}+\xi_{22}}-\alpha(t)^2e^{\xi_{12}+\xi_{21}})}
{(\frac{1}{\lambda}+e^{\xi_{11}})(\frac{1}{\mu}+e^{\xi_{22}})+\alpha(t)^2e^{\xi_{12}+\xi_{21}}},\\
&\Phi=-\dot{\alpha}(t)\left(\begin{array}{cc}
\alpha(t)^2\frac{\frac{\mu}{\lambda}e^{\xi_{12}+\xi_{21}}+(\mu-\lambda)
e^{\xi_{11}+\xi_{12}+\xi_{21}}}
{(\frac{1}{\lambda}+e^{\xi_{11}})(\frac{1}{\mu}+e^{\xi_{22}})+\alpha(t)^2e^{\xi_{12}+\xi_{21}}}
&
-\alpha(t)e^{\xi_{12}}\frac{(\mu-\lambda)e^{\xi_{11}+\xi_{22}}+(1-\frac{\lambda}{\mu})e^{\xi_{11}}+\frac{\mu}{\lambda}e^{\xi_{22}}+\frac{1}{\lambda}}
{(\frac{1}{\lambda}+e^{\xi_{11}})(\frac{1}{\mu}+e^{\xi_{22}})+\alpha(t)^2e^{\xi_{12}+\xi_{21}}}\\
\alpha(t)e^{\xi_{21}}\frac{(\lambda-\mu)e^{\xi_{11}+\xi_{22}}+(1-\frac{\lambda}{\mu})e^{\xi_{22}}+\frac{\lambda}{\mu}e^{\xi_{11}}+\frac{1}{\mu}}
{(\frac{1}{\lambda}+e^{\xi_{11}})(\frac{1}{\mu}+e^{\xi_{22}})+\alpha(t)^2e^{\xi_{12}+\xi_{21}}}
&
\alpha(t)^2\frac{\frac{\lambda}{\mu}e^{\xi_{21}+\xi_{12}}+(\lambda-\mu)
e^{\xi_{21}+\xi_{22}+\xi_{12}}}
{(\frac{1}{\lambda}+e^{\xi_{11}})(\frac{1}{\mu}+e^{\xi_{22}})+\alpha(t)^2e^{\xi_{12}+\xi_{21}}}\\\end{array}\right),\\
&\Psi=\left(\begin{array}{cc} \frac{\frac{1}{\mu}+e^{\xi_{22}}}
{(\frac{1}{\lambda}+e^{\xi_{11}})(\frac{1}{\mu}+e^{\xi_{22}})+\alpha(t)^2e^{\xi_{12}+\xi_{21}}}
& \frac{-\alpha(t)e^{\xi_{21}}}
{(\frac{1}{\lambda}+e^{\xi_{11}})(\frac{1}{\mu}+e^{\xi_{22}})+\alpha(t)^2e^{\xi_{12}+\xi_{21}}}\\
\frac{\alpha(t)e^{\xi_{12}}}
{(\frac{1}{\lambda}+e^{\xi_{11}})(\frac{1}{\mu}+e^{\xi_{22}})+\alpha(t)^2e^{\xi_{12}+\xi_{21}}}
& \frac{\frac{1}{\lambda}+e^{\xi_{11}}}
{(\frac{1}{\lambda}+e^{\xi_{11}})(\frac{1}{\mu}+e^{\xi_{22}})+\alpha(t)^2e^{\xi_{12}+\xi_{21}}}\\\end{array}\right).
\end{align}

The one-soliton solution of the second type of $(2+1)$-AKNS equation
with a self-consitent source (2.24) can be constructed from
$h=f+\alpha(y) g=\left(\begin{array}{cc}\frac{1}{\lambda}+e^{\lambda
x+\lambda y+\lambda^2t} &
-\alpha(y)e^{\mu x+\mu y+\mu^2t}\\
\alpha(y)e^{\lambda x-\lambda y-\lambda^2t} & \frac{1}{\mu}+e^{\mu
x-\mu
y-\mu^2t}\\\end{array}\right):=\left(\begin{array}{cc}\frac{1}{\lambda}+e^{\xi_{11}}
& -\alpha(y)e^{\xi_{12}}\\ \alpha(y)e^{\xi_{21}} &
\frac{1}{\mu}+e^{\xi_{22}}\\\end{array}\right)$.

From Theorem 1 and Theorem 2 we have
$a_0=-h_xh^{-1},~a_1=1,~b_0=h^{-1},$ which gives rise to
\begin{eqnarray}
&&W=\partial-h_xh^{-1},~U=2\sigma_3(h_xh^{-1})^{off},~D=2(h_xh^{-1})^{diag},\notag\\
&&\Phi=-\dot{\alpha}(y)(g_x-h_xh^{-1}g),~\Psi^T=h^{-1}.
\end{eqnarray}

So the one-soliton of the second type of (2+1)-AKNS equation with a
self-consistent source (2.24) is
\begin{align}
&q=-2\alpha(y)\frac{\frac{\lambda}{\mu}e^{\xi_{21}}+(\lambda-\mu)e^{\xi_{21}+\xi_{22}}}
{(\frac{1}{\lambda}+e^{\xi_{11}})(\frac{1}{\mu}+e^{\xi_{22}})+\alpha(y)^2e^{\xi_{12}+\xi_{21}}},\\
&r=2\alpha(y)\frac{-\frac{\mu}{\lambda}e^{\xi_{12}}+(\lambda-\mu)e^{\xi_{11}+\xi_{12}}}
{(\frac{1}{\lambda}+e^{\xi_{11}})(\frac{1}{\mu}+e^{\xi_{22}})+\alpha(y)^2e^{\xi_{12}+\xi_{21}}},\\
&v=2\frac{\frac{\lambda}{\mu}e^{\xi_{11}}-\frac{\mu}{\lambda}e^{\xi_{22}}+(\lambda-\mu)(e^{\xi_{11}+\xi_{22}}-\alpha(y)^2e^{\xi_{12}+\xi_{21}})}
{(\frac{1}{\lambda}+e^{\xi_{11}})(\frac{1}{\mu}+e^{\xi_{22}})+\alpha(y)^2e^{\xi_{12}+\xi_{21}}},\\
&\Phi=-\dot{\alpha}(y)\left(\begin{array}{cc}
\alpha(t)^2\frac{\frac{\mu}{\lambda}e^{\xi_{12}+\xi_{21}}+(\mu-\lambda)
e^{\xi_{11}+\xi_{12}+\xi_{21}}}
{(\frac{1}{\lambda}+e^{\xi_{11}})(\frac{1}{\mu}+e^{\xi_{22}})+\alpha(y)^2e^{\xi_{12}+\xi_{21}}}
&
-\alpha(y)e^{\xi_{12}}\frac{(\mu-\lambda)e^{\xi_{11}+\xi_{22}}+(1-\frac{\lambda}{\mu})e^{\xi_{11}}+\frac{\mu}{\lambda}e^{\xi_{22}}+\frac{1}{\lambda}}
{(\frac{1}{\lambda}+e^{\xi_{11}})(\frac{1}{\mu}+e^{\xi_{22}})+\alpha(y)^2e^{\xi_{12}+\xi_{21}}}\\
\alpha(y)e^{\xi_{21}}\frac{(\lambda-\mu)e^{\xi_{11}+\xi_{22}}+(1-\frac{\lambda}{\mu})e^{\xi_{22}}+\frac{\lambda}{\mu}e^{\xi_{11}}+\frac{1}{\mu}}
{(\frac{1}{\lambda}+e^{\xi_{11}})(\frac{1}{\mu}+e^{\xi_{22}})+\alpha(y)^2e^{\xi_{12}+\xi_{21}}}
&
\alpha(y)^2\frac{\frac{\lambda}{\mu}e^{\xi_{21}+\xi_{12}}+(\lambda-\mu)
e^{\xi_{21}+\xi_{22}+\xi_{12}}}
{(\frac{1}{\lambda}+e^{\xi_{11}})(\frac{1}{\mu}+e^{\xi_{22}})+\alpha(y)^2e^{\xi_{12}+\xi_{21}}}\\\end{array}\right),\\
&\Psi=\left(\begin{array}{cc} \frac{\frac{1}{\mu}+e^{\xi_{22}}}
{(\frac{1}{\lambda}+e^{\xi_{11}})(\frac{1}{\mu}+e^{\xi_{22}})+\alpha(y)^2e^{\xi_{12}+\xi_{21}}}
& \frac{-\alpha(y)e^{\xi_{21}}}
{(\frac{1}{\lambda}+e^{\xi_{11}})(\frac{1}{\mu}+e^{\xi_{22}})+\alpha(y)^2e^{\xi_{12}+\xi_{21}}}\\
\frac{\alpha(y)e^{\xi_{12}}}
{(\frac{1}{\lambda}+e^{\xi_{11}})(\frac{1}{\mu}+e^{\xi_{22}})+\alpha(y)^2e^{\xi_{12}+\xi_{21}}}
& \frac{\frac{1}{\lambda}+e^{\xi_{11}}}
{(\frac{1}{\lambda}+e^{\xi_{11}})(\frac{1}{\mu}+e^{\xi_{22}})+\alpha(y)^2e^{\xi_{12}+\xi_{21}}}\\\end{array}\right).
\end{align}

The one-soliton solution of the first type of the $(2+1)$-DS
equation with a self-consistent source (2.26) can be constructed
from
$$h=f+\alpha(t)
g=\left(\begin{array}{cc}\frac{1}{\lambda}+e^{\lambda x+\lambda
y+i\lambda^2t} &
-\alpha(t)e^{-\lambda x-\lambda y+i\lambda^2t}\\
\alpha(t)e^{\lambda x-\lambda y-i\lambda^2t} &
\frac{1}{\lambda}+e^{-\lambda x+\lambda
y-i\lambda^2t}\\\end{array}\right):=\left(\begin{array}{cc}\frac{1}{\lambda}+e^{\eta_{11}}
& -\alpha(t)e^{\eta_{12}}\\\alpha(t)e^{\eta_{21}} &
\frac{1}{\lambda}+e^{\eta_{22}}\\\end{array}\right).$$ where
$\lambda>0$.

We find that the one-soliton solution of the first type of the
$(2+1)$-DS equation with a self-consistent source is
\begin{align} &r=2\alpha(t)\frac{e^{\eta_{12}}+2\lambda
e^{\eta_{11}+\eta_{12}}}
{(\frac{1}{\lambda}+e^{\eta_{11}})(\frac{1}{\lambda}+e^{\eta_{22}})+\alpha(t)^2e^{\eta_{12}+\eta_{21}}},\\
&v=2\frac{e^{\eta_{22}}+e^{\eta_{11}}+2\lambda(e^{\eta_{11}+\eta_{22}}-\alpha(t)^2e^{\eta_{12}+\eta_{21}})}
{(\frac{1}{\lambda}+e^{\eta_{11}})(\frac{1}{\lambda}+e^{\eta_{22}})+\alpha(t)^2e^{\eta_{12}+\eta_{21}}},\\
&\Phi=-\dot{\alpha}(t)\left(\begin{array}{cc}
-\alpha(t)\frac{e^{\eta_{12}+\eta_{21}}+2\lambda
e^{\eta_{11}+\eta_{12}+\eta_{21}}}
{(\frac{1}{\lambda}+e^{\eta_{11}})(\frac{1}{\lambda}+e^{\eta_{22}})+\alpha(t)^2e^{\eta_{12}+\eta_{21}}}
& \alpha(t)e^{\eta_{12}}\frac{2\lambda
e^{\eta_{11}+\eta_{22}}+2e^{\eta_{11}}+e^{\eta_{22}}+\frac{1}{\lambda}}
{(\frac{1}{\lambda}+e^{\eta_{11}})(\frac{1}{\lambda}+e^{\eta_{22}})+\alpha(t)^2e^{\eta_{12}+\eta_{21}}}\\
\alpha(t)e^{\eta_{21}}\frac{2\lambda
e^{\eta_{11}+\eta_{22}}+2e^{\eta_{22}}+e^{\eta_{11}}+\frac{1}{\lambda}}
{(\frac{1}{\lambda}+e^{\eta_{11}})(\frac{1}{\lambda}+e^{\eta_{22}})+\alpha(t)^2e^{\eta_{12}+\eta_{21}}}
& \alpha(t)\frac{e^{\eta_{21}+\eta_{12}}+2\lambda
e^{\eta_{21}+\eta_{22}+\eta_{12}}}
{(\frac{1}{\lambda}+e^{\eta_{11}})(\frac{1}{\lambda}+e^{\eta_{22}})+\alpha(t)^2e^{\eta_{12}+\eta_{21}}}\\\end{array}\right),\\
&\Psi=\left(\begin{array}{cc} \frac{\frac{1}{\lambda}+e^{\eta_{22}}}
{(\frac{1}{\lambda}+e^{\eta_{11}})(\frac{1}{\lambda}+e^{\eta_{22}})+\alpha(t)^2e^{\eta_{12}+\eta_{21}}}
& \frac{-\alpha(t)e^{\eta_{21}}}
{(\frac{1}{\lambda}+e^{\eta_{11}})(\frac{1}{\lambda}+e^{\eta_{22}})+\alpha(t)^2e^{\eta_{12}+\eta_{21}}}\\
\frac{\alpha(t)e^{\eta_{12}}}
{(\frac{1}{\lambda}+e^{\eta_{11}})(\frac{1}{\lambda}+e^{\eta_{22}})+\alpha(t)^2e^{\eta_{12}+\eta_{21}}}
& \frac{\frac{1}{\lambda}+e^{\eta_{11}}}
{(\frac{1}{\lambda}+e^{\eta_{11}})(\frac{1}{\lambda}+e^{\eta_{22}})+\alpha(t)^2e^{\eta_{12}+\eta_{21}}}\\\end{array}\right).
\end{align}

The one-soliton solution of the second type of the $(2+1)$-DS
equation with a self-consistent source (2.28) can be constructed
from
$$h=f+\alpha(y)
g=\left(\begin{array}{cc}\frac{1}{\lambda}+e^{\lambda x+\lambda
y+i\lambda^2t} &
-\alpha(y)e^{-\lambda x-\lambda y+i\lambda^2t}\\
\alpha(y)e^{\lambda x-\lambda y-i\lambda^2t} &
\frac{1}{\lambda}+e^{-\lambda x+\lambda
y-i\lambda^2t}\\\end{array}\right):=\left(\begin{array}{cc}\frac{1}{\lambda}+e^{\eta_{11}}
& -\alpha(y)e^{\eta_{12}}\\\alpha(y)e^{\eta_{21}} &
\frac{1}{\lambda}+e^{\eta_{22}}\\\end{array}\right).$$ where
$\lambda>0$.

We find that the one-soliton solution of the second type of the
$(2+1)$-DS equation with a self-consistent source is
\begin{align}
&r=2\alpha(y)\frac{e^{\eta_{12}}+2\lambda e^{\eta_{11}+\eta_{12}}}
{(\frac{1}{\lambda}+e^{\eta_{11}})(\frac{1}{\lambda}+e^{\eta_{22}})+\alpha(y)^2e^{\eta_{12}+\eta_{21}}},\\
&v=2\frac{e^{\eta_{22}}+e^{\eta_{11}}+2\lambda(e^{\eta_{11}+\eta_{22}}-\alpha(y)^2e^{\eta_{12}+\eta_{21}})}
{(\frac{1}{\lambda}+e^{\eta_{11}})(\frac{1}{\lambda}+e^{\eta_{22}})+\alpha(y)^2e^{\eta_{12}+\eta_{21}}},\\
&\Phi=-\dot{\alpha}(y)\left(\begin{array}{cc}
-\alpha(y)\frac{e^{\eta_{12}+\eta_{21}}+2\lambda
e^{\eta_{11}+\eta_{12}+\eta_{21}}}
{(\frac{1}{\lambda}+e^{\eta_{11}})(\frac{1}{\lambda}+e^{\eta_{22}})+\alpha(y)^2e^{\eta_{12}+\eta_{21}}}
& \alpha(y)e^{\eta_{12}}\frac{2\lambda
e^{\eta_{11}+\eta_{22}}+2e^{\eta_{11}}+e^{\eta_{22}}+\frac{1}{\lambda}}
{(\frac{1}{\lambda}+e^{\eta_{11}})(\frac{1}{\lambda}+e^{\eta_{22}})+\alpha(y)^2e^{\eta_{12}+\eta_{21}}}\\
\alpha(y)e^{\eta_{21}}\frac{2\lambda
e^{\eta_{11}+\eta_{22}}+2e^{\eta_{22}}+e^{\eta_{11}}+\frac{1}{\lambda}}
{(\frac{1}{\lambda}+e^{\eta_{11}})(\frac{1}{\lambda}+e^{\eta_{22}})+\alpha(y)^2e^{\eta_{12}+\eta_{21}}}
& \alpha(y)\frac{e^{\eta_{21}+\eta_{12}}+2\lambda
e^{\eta_{21}+\eta_{22}+\eta_{12}}}
{(\frac{1}{\lambda}+e^{\eta_{11}})(\frac{1}{\lambda}+e^{\eta_{22}})+\alpha(y)^2e^{\eta_{12}+\eta_{21}}}\\\end{array}\right),\\
&\Psi=\left(\begin{array}{cc} \frac{\frac{1}{\lambda}+e^{\eta_{22}}}
{(\frac{1}{\lambda}+e^{\eta_{11}})(\frac{1}{\lambda}+e^{\eta_{22}})+\alpha(y)^2e^{\eta_{12}+\eta_{21}}}
& \frac{-\alpha(y)e^{\eta_{21}}}
{(\frac{1}{\lambda}+e^{\eta_{11}})(\frac{1}{\lambda}+e^{\eta_{22}})+\alpha(y)^2e^{\eta_{12}+\eta_{21}}}\\
\frac{\alpha(y)e^{\eta_{12}}}
{(\frac{1}{\lambda}+e^{\eta_{11}})(\frac{1}{\lambda}+e^{\eta_{22}})+\alpha(y)^2e^{\eta_{12}+\eta_{21}}}
& \frac{\frac{1}{\lambda}+e^{\eta_{11}}}
{(\frac{1}{\lambda}+e^{\eta_{11}})(\frac{1}{\lambda}+e^{\eta_{22}})+\alpha(y)^2e^{\eta_{12}+\eta_{21}}}\\\end{array}\right).
\end{align}

\section{Conclusion}
\setcounter{equation}{0} We extend the matrix KP hierarchy by
introducing a new $\tau_B$ flow and adding eigenfunctions and
adjoint eigenfunctions as new components. The zero curvature
equation and Lax representation for the extended matrix KP hierarchy
and its $t_A$-reduction and $\tau_B$-reduction are presented. The
extended matrix KP hierarchy and its two reductions provide an
unified way to find two types of (2+1)-dimensional and
(1+1)-dimensional AKNS equation (and DS equation) with
self-consistent sources. With the combination of dressing method and
the method of variation of constants, we propose a generalized
dressing method to solve the extended matrix KP hierarchy and obtain
some of its solutions. The soliton solution of two types of 2+1 AKNS
equation with self-consistent sources and two types of DS equation
with self-consistent sources are studied.

\section*{Acknowledgements}
This work was supported by the National Basic Research Program of
China (973 program) \\(2007CB814800) and the National Science
Foundation of China (Grant no 10801083, 10901090).


\begin{thebibliography}{99}
\bibitem {1} Sato M, RIMS Kokyuroku, {\bf 439}, 30 (1981).
\bibitem {2} Date E, Jimbo M, Kashiwara M and Miwa T, J. phys. Soc. Jpn., {\bf
50},
3806 (1981).
\bibitem {3} Oevel W,  Phys. A, {\bf 195}, 533 (1993).
\bibitem {4} Kundu A and Strampp W, J. Math. Phys., {\bf 36},
4192 (1995).
\bibitem {5} Mel'nikov V K, Lett. Math. Phys., {\bf 7},
129 (1983).
\bibitem {6} Mel'nikov V K, Commun. Math. Phys., {\bf 112},
639 (1987).
\bibitem {7} Mel'nikov V K, Commun. Math. Phys., {\bf 126},
201 (1989).
\bibitem {8} Chen Y, J. Math. Phys., {\bf 33}, 3774
(1992).
\bibitem {9} Xiao T and Zeng Y B, J. Phys. A, {\bf 37},
7143 (2004).
\bibitem {10} Sakhonovich A L, J. Phys. A: Math. Gen., {\bf{36}},
5023 (2003).
\bibitem {11} Schiebold C, Glasgow. Math. J., {\bf 51A},
147 (2009).
\bibitem {12} Zenchuk A I and Santini P M, J. Phys. A: Math. Theor., {\bf 41},
185209 (2008).
\bibitem {13} Liu X J, Zeng Y B and Lin R L, Phys. Lett. A, {\bf
372},
3819 (2008).
\bibitem {14} Liu X J, Lin R L, Jin B and Zeng Y B, J. Math. Phys., {\bf
50},
053506 (2009).
\bibitem {15} Hu J, Wang H Y and Tam H W, J. Math. Phys., {\bf 49},
013506 (2008).
\bibitem {16} Shen S and Jiang L, J. Comp. Appl. Math., {\bf 233},
585 (2009).
\bibitem {17} Dickey L A, Soliton Equations and Hamiltonian
Systems, Advanced Series in Mathematical Physics, 2nd ed (World
Scientific, River Edge, NJ, Vol. 26. (2003).

\end{thebibliography}
\end{document}